\documentclass[12pt]{article}
\usepackage[dvips]{graphicx}
\usepackage{epsfig}
\usepackage{cite}
\usepackage{latexsym,amsmath,amsfonts,amssymb}
\usepackage{mathtools}
\usepackage{slashed}
\usepackage{blkarray}
\usepackage[all]{xy}
\usepackage[makeroom]{cancel}
\usepackage[latin1]{inputenc}
\usepackage[american]{babel}
\usepackage[dvips]{graphicx}
\usepackage{bbm}
\usepackage{color}
\usepackage[unicode]{hyperref}
\def\6{\partial}

\newcommand{\be}{\begin{equation}}
\newcommand{\ee}{\end{equation}}
\newcommand{\beq}{\begin{equation}}
\newcommand{\eeq}{\end{equation}}
\newcommand{\bea}{\begin{eqnarray}}
\newcommand{\eea}{\end{eqnarray}}
\newcommand{\nn}{\nonumber \\}

\newcommand{\ba}{\begin{eqnarray}}
\newcommand{\ea}{\end{eqnarray}}

\newcommand{\beqs}{\begin{eqnarray}}
\newcommand{\eeqs}{\end{eqnarray}}
\newcommand{\bal}{\begin{aligned}}
\newcommand{\eal}{\end{aligned}}

\def\lbldef#1#2{\expandafter\gdef\csname #1\endcsname {#2}}

\def\href#1#2{#2}

\newcommand{\ber}{\begin{eqnarray}}
\newcommand{\eer}{\end{eqnarray}}

\newcommand{\beqar}{\begin{eqnarray}}

\newcommand{\eeqar}{\end{eqnarray}}

\newcommand{\dsl}
   {\kern.06em\hbox{\raise.15ex\hbox{$/$}\kern-.56em\hbox{$\partial$}}}

\newcommand{\eeqarr}{\end{eqnarray}}
\newcommand{\ZZ}{{\rm \kern 0.275em Z \kern -0.92em Z}\;}

 \def\vol{{\hbox{\rm vol}}}

\def\CC{{\mathchoice
{\rm C\mkern-8mu\vrule height1.45ex depth-.05ex
width.05em\mkern9mu\kern-.05em}
{\rm C\mkern-8mu\vrule height1.45ex depth-.05ex
width.05em\mkern9mu\kern-.05em}
{\rm C\mkern-8mu\vrule height1ex depth-.07ex
width.035em\mkern9mu\kern-.035em}
{\rm C\mkern-8mu\vrule height.65ex depth-.1ex
width.025em\mkern8mu\kern-.025em}}}

\def\RR{{\rm I\kern-1.6pt {\rm R}}}

\def\ZZ{{\rm Z}\kern-3.8pt {\rm Z} \kern2pt}
\def\IB{\relax{\rm I\kern-.18em B}}
\def\ID{\relax{\rm I\kern-.18em D}}
\def\II{\relax{\rm I\kern-.18em I}}
\def\IP{\relax{\rm I\kern-.18em P}}

\newcommand{\bear}{\begin{eqnarray}}
\newcommand{\eear}{\end{eqnarray}}

\def\e{\epsilon}           

\def\6{\partial}

\newfont{\namefont}{cmr10}
\newfont{\addfont}{cmti7 scaled 1440}
\newfont{\boldmathfont}{cmbx10}
\newfont{\headfontb}{cmbx10 scaled 1728}

\newcommand{\dd}{\textrm{d}}
\newcommand{\DD}{\textrm{D}}

\allowdisplaybreaks[1]

\numberwithin{equation}{section}

\newcommand{\Tr}{\mbox{Tr}}    

\begin{document}

\begin{titlepage}

\vfill
\begin{flushright}
\end{flushright}

\vfill

\begin{center}
   \baselineskip=16pt
   {\Large \bf Ehlers as EM duality in the double copy}
   \vskip 2cm
   A. Banerjee$^{a}$, E. \'O Colg\'ain$^{a, b}$, J. A. Rosabal$^a$, H. Yavartanoo$^{c}$
          \vskip .6cm
             \begin{small}
               \textit{$^a$ Asia Pacific Center for Theoretical Physics, Postech, Pohang 37673, Korea}
               
               \vspace{3mm} 
               
               \textit{$^b$ Department of Physics, Postech, Pohang 37673, Korea}
               
               \vspace{3mm} 
               
               \textit{$^c$ CAS Key Laboratory of Theoretical Physics, Institute of Theoretical Physics, \\ Chinese Academy of Sciences, Beijing 100190, China}
  
             \end{small}
\end{center}

\vfill \begin{center} \textbf{Abstract}\end{center} \begin{quote}
Given a solution to 4D Einstein gravity with an isometry direction, it is known that the equations of motion are identical to those of a 3D $\sigma$-model with target space geometry $SU(1,1)/U(1)$. Thus, any transformation by $SU(1, 1) \cong SL(2,\mathbb{R})$ is a symmetry for the action and allows one to generate new solutions in 4D. Here we clarify and extend recent work on electromagnetic (EM) duality in the context of the classical double copy. In particular, for pure gravity, we identify an explicit map between the Maxwell field of the single copy and the scalars in the target space, allowing us to identify the $U(1) \subset SL(2, \mathbb{R})$ symmetry dual to EM duality in the single copy. Moreover, we extend the analysis to Einstein-Maxwell theory, where we highlight the role of Ehlers-Harrison transformations and,
{for spherically symmetric charged black hole solutions, we interpret the equations of motion as a truncation of the putative single copy for Einstein-Yang-Mills theory.}

\end{quote} \vfill

\end{titlepage}

\section{Introduction}
The classical double copy is an intriguing connection between gravity and gauge theory \cite{Monteiro:2014cda}, which has been motivated from a relationship between perturbative scattering amplitudes in gauge theory and gravity \cite{Bern:2008qj, Bern:2010ue, Bern:2010yg} \footnote{See the recent review \cite{Bern:2019prr} and references therein for a wider perspective on this.}. In its simplest form, the central observation is that solutions to Einstein gravity, or ``the double copy", can be mapped to solutions of Maxwell's equations \footnote{Here Maxwell may be viewed as a linearisation of the Yang-Mills theory that features in perturbative statements at the level of scattering amplitudes.}, or ``the single copy", through a Kerr-Schild (KS) decomposition of the spacetime. Interestingly, in contrast to Kaluza-Klein dimensional reduction, the KS ansatz maintains dimensionality. More concretely, one considers the spacetime metric  
\be
\label{KS}
g_{\mu \nu} = \eta_{\mu \nu} + \phi \, k_{\mu} k_{\nu},
\ee
where $\eta_{\mu \nu}$ denotes the metric of flat spacetime, $\phi$ is a scalar and $k$ is a null vector, $k_{\mu} k^{\mu} = 0$, satisfying the geodesic equation $k^{\rho} \partial_{\rho} k_{\mu} = 0$. The Maxwell gauge field $A$ arises from the identification $A  = \phi \, k$. See \cite{Sabharwal:2019ngs, Anastasiou:2014qba,Borsten:2015pla, Ridgway:2015fdl, Luna:2016due, Carrillo-Gonzalez:2017iyj, Anastasiou:2016csv,Anastasiou:2017nsz,Cardoso:2016ngt, Cardoso:2016amd, Borsten:2017jpt,Anastasiou:2017taf, Anastasiou:2018rd, Anastasiou:2018rdx, Gurses:2018ckx,  LopesCardoso:2018xes, Goldberger:2019xef, Lee:2018gxc, Cho:2019ype, Kim:2019jwm} for related work in this direction. 

In this double copy formalism the Schwarzschild solution corresponds to a Maxwell field with an electric charge \cite{Monteiro:2014cda}, while the Taub-NUT solution possesses a magnetic charge \cite{Luna:2015paa}. Subsequently,  the single copy of the Eguchi-Hanson instanton has been shown to map to a self-dual Maxwell field \cite{Berman:2018hwd}. With both electric and magnetic charges present, this raises the question whether there is a gravity analogue of electromagnetic (EM) duality, namely a rotation of the field strength $F = \dd A$ into $* F$ that honours the Maxwell equations of motion. This was answered in the affirmative in two recent papers. In the first a complex transformation in the gravity is mapped to a complexified BMS supertranslation \cite{Huang:2019cja}, while in the second \cite{Alawadhi:2019urr} a class of real transformations due to Ehlers \cite{Ehlers:1961zza} (also Geroch\cite{Geroch:1970nt}) are exploited. 

One goal of this work is to clarify comments in the latter paper. As we explain in the following section, the magic of Ehlers transformations is that given 4D \textit{pure} gravity with a $U(1)$ isometry direction, the equations of motion are identical to a 3D $\sigma$-model with a target space $H^2$. Being maximally symmetric, the hyperbolic space $H^2$ possesses an isometry group $SU(1,1) \cong SL(2, \mathbb{R})$ that rotates the scalars, but importantly leaves the 3D effective action, and therefore the equations of motion, invariant. Of these three $SL(2, \mathbb{R})$ transformations, one corresponds to a trivial shift that is pure gauge, a second to a constant rescaling of the Killing vector of the $U(1)$ isometry direction, while it is the third ``Ehlers transformation" \footnote{Replacing $H^2$ with $AdS_2$, Ehlers is the analogue of a special conformal transformation, while the shift is a translation and the scale symmetry is dilatation.} that is non-trivial in 4D. In section \ref{sec:ehlers}, we identify a linear combination of these transformations as the appropriate $U(1)$ transformation that is EM duality in the single copy. Our lower-dimensional approach here should be contrasted with \cite{Alawadhi:2019urr}, where due to the fact that one is working in 4D, the simplicity of the mapping between the single and double copy is obscured. In short, Ehlers is simpler in 3D. 

Concretely, in section \ref{sec:ehlers} we rewrite the KS ansatz in a way appropriate for dimensional reduction on a timelike direction. In essence, we are combining the classical double copy in pure gravity with Kaluza-Klein reduction and the beauty of this approach is that a manifest $U(1)$ symmetry in 4D leads to an $SL(2, \mathbb{R})$ symmetry in the lower-dimensional theory. Through this process, we show how the electric and magnetic Maxwell field strengths of the single copy are related to the derivatives of the scalars of the 3D $\sigma$-model in the double copy, thereby providing a succinct way to understand observations made in \cite{Alawadhi:2019urr}. This map between the scalars parametrising $H^2$ and the Maxwell fluxes allows us to define electric and magnetic Maxwell charges at the level of the 3D effective theory, which transform accordingly. Our construction can be extended to a double KS ansatz, which makes us believe that it holds for all stationary spacetimes admitting a KS decomposition.

The generalisation from Ehlers transformation to Ehlers-Harrison transformations \cite{Harrison} in 4D Einstein-Maxwell theory is immediate. Given the richer field content in 4D, the symmetries of the target spacetime of the 3D $\sigma$-model are enhanced from $SU(1,1) \rightarrow SU(2,1)$  \cite{Maison:1979kx, Kinnersley:1977pg, Kinnersley:1977ph, Galtsov:1995mb}. {Nevertheless, the interpretation of these enlarged symmetries in the double copy is unclear. More precisely,} despite a host of perturbative results at the level of scattering amplitudes \cite{Chiodaroli:2014xia, Chiodaroli:2015rdg, Chiodaroli:2016jqw} and radiation \cite{Goldberger:2016iau, Chester:2017vcz} suggesting that {Einstein-Yang-Mills can be formulated as the double copy of pure
Yang-Mills and Yang-Mills coupled to a bi-adjoint scalar with cubic potential}, it is currently not known how to define the classical double copy for black hole spacetimes where the metric, in particular the $g_{tt}$ term, scales with the radial direction $r$ as $r^{-n}, n > 1$ \footnote{{See \cite{Cho:2019ype} for black hole solutions in Einstein-Maxwell-dilaton theory and their interpretation in terms of the single copy. The examples given are close cousins of the Schwarzschild solution in the sense that the $g_{tt}$ term is linear in $1/r$.}}. This raises an interesting puzzle concerning the single copy interpretation of charged black holes,  especially black holes that are related to the Schwarzschild solution through Ehlers-Harrison transformations. In the latter part of this work (section \ref{sec:harrison}), we identify the relevant equations of motions for a class of spherically symmetric charged black holes and show that similar equations may be found from a truncation of the putative single copy.

{Concretely}, we show through a generalised KS decomposition \cite{Vaidya:1947zz}, that the equations of motion for the Maxwell fields reduce to the same equations of motion evaluated on flat spacetime. Somewhat surprisingly, this implies that the Maxwell field strength with a KS ansatz is always a harmonic two-form on flat spacetime! Secondly, we observe that the Harrison transformation, which turns on electric and/or magnetic charges in black holes, is new to the double copy literature. {It should be noted that Harrison transformations generate $r^{-2}$ terms in the metric from $r^{-1}$ expressions, thereby taking one outside of the current classical double copy prescription. \footnote{We note that all 4D black holes  at order $G$ and charge $e^2$ can be obtained from minimal coupling via tree-level and one-loop triangle leading singularities \cite{Moynihan:2019bor}. Interestingly, the Harrison transformation is a classical transformation that appears to rotate tree level and one-loop triangle diagrams into each other.}} Thirdly, we observe the same KS decomposition allows us to interpret an additional equation as that of a {truncation of the biadjoint scalar equation \cite{White:2016jzc}, itself the expected single copy for Yang-Mills theory. This provides potentially the first hint of the biadjoint scalar equation beyond linear order in the classical double copy approach}. Finally, we illustrate how the Ehlers transformation once again plays the counterpart of EM duality in this extended setting.

\section{Ehlers and double copy} 
\label{sec:ehlers}
Here we follow the treatment described in appendix of \cite{Bakhmatov:2019dow} for pure gravity in 4D, which serves as a warm-up for the later extension to Einstein-Maxwell theory. Consider a 4D spacetime with a Killing vector $\partial_t$, which we will assume is in the temporal direction. The most general metric consistent with this $U(1)$ symmetry is 
\be
\label{ehlers}
\dd s^2 = - V ( \dd t + \mathcal{A})^2 + V^{-1} \gamma_{mn} \dd x^m \dd x^n, 
\ee
where $V$ is a scalar and $\mathcal{A}$ is a vector on the transverse 3D space with metric $\gamma_{mn}$, $m, n = 1, 2, 3$. We have rescaled the internal space judiciously so as to arrive later in Einstein frame in 3D. Now, let us demand that this is a vacuum solution to Einstein gravity, so that it satisfies the equation 
\be
R_{\mu \nu} = 0.  
\ee
The joy of this set-up is that the equation mixing the temporal and spatial directions reduces to 
\be
\label{eq0}
\dd (V^2 *_3 \mathcal{F}) = 0, 
\ee
where $\mathcal{F}$ is the field strength corresponding to the vector field, $\mathcal{F} = \dd \mathcal{A}$. Now comes the magic. Locally, one can replace the above equation with 
\be
\label{key}
V^2 *_3 \mathcal{F} = \dd \chi, 
\ee
where we have taken the opportunity to introduce a second scalar. The fact that we can do this is essentially down to dimensionality: \textit{in 3D vectors are dual to scalars}. Gathering the remaining equations of motion together, it can be shown that the equations of motion follow from varying the following 3D action 
\be
\label{action}
\mathcal{L} = \sqrt{\gamma} \left( R - \frac{1}{2} \frac{\partial_m V \partial^m V + \partial_m \chi \partial^m \chi }{V^2} \right). 
\ee
From the action it is evident that there is a hyperbolic target space $H^2$. Being maximally symmetric, it permits 3 Killing directions. To make these symmetries manifest, it is best to switch to the complex scalar 
\be
\tau = \chi + i V, 
\ee
which allows us to rewrite the metric on the hyperbolic space as
\be
\dd s^2 (H^2) = \frac{\dd V^2 + \dd \chi^2}{V^2} = \frac{\dd \tau \dd \bar{\tau}}{\textrm{Im}(\tau)^2}. 
\ee
It is now an easy task to confirm that the 2D metric, and thus the 3D action, is invariant under the $SL(2, \mathbb{R})$ transformation
\be
\label{sl2R}
\tau \rightarrow \tau' = \frac{a \tau + b}{c \tau + d}, \quad a d - bc = 1, \quad a, b, c, d, \in \mathbb{R}. 
\ee
We believe that this is the simplest and most elegant way to present the class of transformations attributable to Ehlers/Geroch \cite{Ehlers:1961zza, Geroch:1970nt} \footnote{The reader is welcome to compare with sections 3 and 5 of the recent paper \cite{Alawadhi:2019urr}, where the same transformation is discussed in 4D and the underlying simplicity is lost. It is helpful to note that $\tau = i \sigma$. }. In appendix \ref{sec:coset} we provide a coset description for the same transformation. 

Just so we are all on the same page, some comments are in order. First, the $SL(2, \mathbb{R})$ clearly rotates the scalars in the action, but does not affect the 3D Ricci scalar. For this reason, the 3D space with metric $\gamma_{mn}$ is indeed \textit{invariant}. Second, although we appear to have three free parameters, the freedom to rescale the Killing vector by a constant and the freedom to shift $\chi$ by a constant removes two of these parameters. In effect, if one is interested in generating new inequivalent solutions in 4D, one has only one parameter to play with. To see this explicitly, it is worth observing that the following matrix corresponds to transformations that are either pure gauge or can be removed by rescaling \cite{Geroch:1970nt}: 
\be
\label{puregauge}
\left( \begin{array}{cc} a & b \\ 0 & a^{-1} \end{array} \right) \subset SL(2, \mathbb{R}). 
\ee

Interestingly, as explicitly highlighted in \cite{Bakhmatov:2019dow}, the same $SL(2, \mathbb{R}$) symmetry is at the heart of Lunin \& Maldacena's TsT transformations \cite{Lunin:2005jy}, and there one finds only one parameter, in line with expectations. We explicitly check in appendix \ref{sec:gensl2} that the most general $SL(2, \mathbb{R})$ transformation applied to the Schwarzschild solution leads to the Taub-NUT solution, i. e. in addition to the mass, only one additional charge is generated. This further confirms that there is only one relevant parameter.

\subsection{Kerr-Schild}
Now comes a key point of this work. To fully understand the Ehlers transformation in terms of the double copy, one should start with the KS ansatz and identify the scalar $V$ and vector field $\mathcal{A}$ in terms of $\phi$ and the null vector $k$. The only problem is that nowhere in the KS ansatz is a Killing direction specified, so we will have to put one in by hand. Luckily for us, for stationary spacetimes, the most general null vector $k$ can be decomposed as 
\be
k = \dd t + \tilde{k}, 
\ee
where $\tilde{k}$ is a spatial vector with unit norm $\tilde{k}_m \tilde{k}^m = 1$. Once this is done, one can easily identify the electric and magnetic part of the Maxwell field, 
\be
F_{\textrm{elec}} = \dd \phi \wedge \dd t, \quad F_{\textrm{mag}} = \dd ( \phi \tilde{k}), 
\ee
where we have opted to use differential forms. Translated into the language of the earlier dimensional reduction, one finds 
\bea
V&=& (1-\phi), \quad \mathcal{A} =  \frac{\phi}{1-\phi} \tilde{k}, \nn
\gamma_{mn} \dd x^m \dd x^n &=& (1 - \phi) \dd \vec{x}^2 + \phi \tilde{k}^2.  
\eea
With this mapping, it is easy to identify the electric Maxwell flux in terms of the derivative of the scalar $V$: 
\be
\label{Fe}
F_{\textrm{elec}} = - \dd V \wedge \dd t. 
\ee
The magnetic Maxwell flux requires a little more work, but in the end takes a simple form. Using the condition $\tilde{k}^n \partial_n \tilde{k}_m = 0$, it is a straightforward calculation (appendix \ref{sec:doubleKS}) to show that 
\be
V^2 *_3 \mathcal{F} = *_3 F_{\textrm{mag}}, 
\ee
where the Hodge duality on the l.h.s. is with respect to the metric $\gamma_{mn}$, whereas on the r.h.s. the metric is $\delta_{mn}$. Since the Maxwell field is assumed to live in flat spacetime, this is in line with expectations. Now, returning to the key point in the Ehlers transformation, where the vector is replaced by a scalar, we can write 
\be
\label{Fm}
*_3 F_{\textrm{mag}} = \dd \chi. 
\ee
Together (\ref{Fe}) and (\ref{Fm}) define an explicit mapping between the Maxwell field strengths in the single copy and the scalars of the 3D effective description of pure gravity, namely the double copy. Note, \textit{this is a general statement that holds for stationary spacetimes admitting a KS description}. This is one of our main results, which generalises statements in \cite{Alawadhi:2019urr} beyond explicit solutions.

\subsection{EM duality}
At this juncture it should be clear that any transformation of the scalars under $SL(2, \mathbb{R})$ is mapped into a transformation in the Maxwell fields. The task remains to identify the precise linear combination of $SL(2, \mathbb{R})$ generators corresponding to EM duality, or more concretely, the following transformation:
\be
\label{EM_duality}
G \rightarrow e^{2 i \beta} G, \quad G \equiv F + i *_4 F, 
\ee
where $\beta$ is a constant parameter, which ensures the equations $\dd F = \dd *_4 F = 0$ hold. 

The search for this constant parameter is made easy by the fact that, as explained earlier, there is only one relevant parameter, so the most general $SL(2, \mathbb{R})$ transformation may be expressed as 
\be
\label{rot}
\left( \begin{array}{cc} a & b \\ c & d \end{array} \right) =  \left( \begin{array}{cc} \cos \beta & \sin \beta \\ -\sin \beta & \cos \beta \end{array} \right). 
\ee
Expanding this transformation for small $\beta$, we see that it is generated by a combination of Ehlers/shift and gives rise to a rescaling, but as explained, the non-trivial solution generating element is coming from the Ehlers transformation. We have opted for this form of transformation as it preserves $V \rightarrow 1$ asymptotically, which is a necessary condition for the spacetime to be asymptotically flat. Then, bearing in mind that $\chi$ can be shifted by a constant, asymptotically one has $V = 1 + {M}/{r} + \dots$ and $\chi = {N}/{r} + \dots$, where $M, N$ denote constant charges (see appendix \ref{sec:gensl2} for the relevant expressions for the Schwarzschild spacetime). The requirement that $V =1$ asymptotically is enough to fix $c^2 + d^2 =1$ and the form (\ref{rot}) follows from the constraint $a d - bc =1$. Ultimately, this guarantees that the charges rotate in the expected manner in line with (\ref{EM_duality}). Therefore, the asymptotic condition $V=1$ is enough to fix (\ref{rot}) uniquely. \footnote{Now, the astute reader will note that Ehlers is inherently a non-linear transformation, which is expected in gravity. This then implies that the Maxwell fields defined in (\ref{Fe}) and (\ref{Fm}) are also transformed non-linearly under (\ref{rot}). Nevertheless, as can be shown for explicit solutions (see subsection \ref{sec:ehlersEM}, appendix \ref{sec:gensl2} or \cite{Alawadhi:2019urr}), such non-linearity can be removed by coordinate transformation. For this reason, we expect that (\ref{rot}) recovers (\ref{EM_duality}) up to a coordinate transformation and that this can be checked on a case-by-case basis.}

Returning to the above transformation (\ref{rot}), we can now comment on some special cases. The choice $\beta = \frac{\pi}{4}$ generates the pure NUT space, while $\beta = \frac{\pi}{2}$ executes the Buchdahl reciprocal transformation \cite{Buchdahl:1959nk}. In contrast to \cite{Alawadhi:2019urr}, there is no need to rescale to the Schwarzschild metric \footnote{This rescaling can be viewed as yet another $SL(2, \mathbb{R}$) transformation where $ d = 1/a$. To make comparison with the Taub-NUT geometry presented in \cite{Momeni:2005uc}, and reproduced in section 3 of \cite{Alawadhi:2019urr}, note that $\sin^2 \beta = \frac{c_1^2}{1+c_1^2}$.} or treat the Buchdahl transformation separately: everything naturally fits into $SL(2, \mathbb{R})$.  It is worth noting that above we have assumed a KS ansatz, but it turns out that the above relations (\ref{Fe}) and (\ref{Fm}) are robust. In appendix \ref{sec:doubleKS} we show that if one replaces a single KS ansatz with the double KS ansatz \footnote{The Plebanski-Demianski family of metrics \cite{Plebanski:1976gy} admits a double KS ansatz once the coordinates are complexified, so this is in principle a large class.}, then the same relations hold. In essence, provided the spacetime admits a (double) KS description, which we should recall is the key assumption in the classical double copy narrative, then we can relate the electric and magnetic fluxes in the single copy to scalars in a 3D $\sigma$-model through (\ref{Fe}) and (\ref{Fm}). It is worth stressing again that $(2.15)$ only holds for KS spacetimes.

Nevertheless, there is an important caveat to our treatment here. It is not guaranteed that Ehlers transformations preserve the KS description and the classification of such solutions is an open problem. In fact, even an Ehlers transformation applied to the Schwarzschild solution does not preserve the single KS description, but the resulting Taub-NUT spacetime admits a more general double KS description \cite{Luna:2015paa}.  For this reason, solutions preserving a strict KS description are expected to be constrained: Schwarzschild is precluded! Of course, if the transformed geometry is not KS, then the classical double copy narrative fails to apply. Thus, only for spacetimes where the Ehlers transformation honours the KS description \footnote{It can be single or double KS.} can one interpret Ehlers in the double copy as EM duality in the single copy. It should be noted that related statements on EM duality in the single copy are very much solution dependent (Schwarzschild $\rightarrow$ Taub-NUT) \cite{Huang:2019cja, Alawadhi:2019urr}, so our analysis here is in principle more general, albeit the classification of Ehlers-invariant KS spacetimes is an open problem we do not address.

Setting this important caveat aside, provided one is careful about the asymptotics, it is possible to define electric and magnetic Maxwell charges in the usual manner: 
\be
Q_{e} = \frac{1}{4 \pi} \int_{S^2} *_3 \dd V, \quad  Q_{m} = \frac{1}{4 \pi} \int_{S^2} *_3 \dd \chi. 
\ee
To the extent of our knowledge these charges are new in the double copy literature, but are not new in general (see for example \cite{Bossard:2009at}, equation (2.9)). This constitutes our second key result. 
We can motivate the charges both physically and mathematically. Physically, the Maxwell fields strengths of the single copy are defined in flat spacetime and we have seen that they are related to the scalars through (\ref{Fe}) and (\ref{Fm}). Moreover, we have removed the temporal direction, so this means we are restricted to the 3D flat metric $\dd s^2 = \dd r^2 + r^2 \dd s^2(S^2)$. Thus, this motivates the integrals on purely physical grounds.

However, we can also rigorously define the asymptotics mathematically. To do so, let us momentarily redefine the scalar $V \rightarrow e^{\phi}$, so that the scalar equations of motion become 
\bea
\dd \left( *_3 \dd \phi  - 2 e^{-2 \phi} \chi *_3 \dd \chi \right)  &=& 0,  \nn
\dd ( e^{-2 \phi} *_3 \dd \chi) &=& 0. 
\eea
Being conserved quantities, the two-forms in the brackets are suitable for defining charges and can be integrated over a closed 2D submanifold. Clearly if $V \rightarrow 1$ ($\phi \rightarrow 0$) and $\chi \rightarrow 0$ asymptotically as $r \rightarrow \infty$, then we recover the charges defined above. As is clear from the 4D metric (\ref{ehlers}), $V=1$ is a necessary condition for the metric to be asymptotically flat, whereas $\chi = 0$ can be easily imposed by exploiting the translational (gauge) symmetry.

Let us return the example considered in \cite{Alawadhi:2019urr} to which we will apply our general one-parameter rotation (\ref{rot}). The data describing the Schwarzschild solution is  
\be
V = \left( 1 - \frac{2M}{r} \right),  
\ee 
and $\chi$ is a constant, so there is no vector field $\mathcal{A}$ in the gravity. Translated into the single copy Maxwell field, the Schwarzschild solution has only an electric flux. In this case we have 
\be
Q_e = 2 M, \quad Q_m = 0. 
\ee 
Performing the $SL(2, \mathbb{R})$ transformation, we generate new scalars and from there we read off the transformed charges, 
\be
Q'_e = 2 M \cos 2 \beta, \quad Q'_m = -2 M \sin 2 \beta. 
\ee

As a further simple example, it is easy to convince oneself that the Buchdahl reciprocal transformation \cite{Buchdahl:1959nk} flips the sign of electric charge. 

\section{Ehlers-Harrison and double copy} 
\label{sec:harrison}
Admittedly, the presentation in the last section has been to the point, but probably a bit quick - we did not provide any details of the dimensional reduction - and some subtleties may still require teasing out. So, in this section we extend the analysis to the Einstein-Maxwell theory in 4D and work with explicit solutions.  The motivation comes from the observation {that charged black holes permit} a \textit{generalised KS} description \footnote{This possibility has already been noticed in \cite{Carrillo-Gonzalez:2017iyj} within the context of the classical double copy.}, a feature we will explain in due course. But before going there, let us explain how the assumption that solutions to Einstein-Maxwell theory in 4D admit a single $U(1)$ Killing vector leads to a hidden $SU(2,1)$ symmetry in 3D. See \cite{Breitenlohner:1987dg} for seminal work in this direction. 

Let us start with the 4D action 
\be
\label{EM}
\mathcal{L}_4 = \sqrt{-g} \left( R - \frac{1}{4} F_{\alpha \beta} F^{\alpha \beta} \right), 
\ee
where $F = \dd A$ is the field strength for a Maxwell gauge field $A$. Now, consider the following ansatz for the spacetime metric and one-form, 
\bea
\dd s^2_4 &=& - V ( \dd t + \mathcal{A})^2 + V^{-1} \gamma_{mn} \dd x^m \dd x^n, \nn
A &=& f ( \dd t + \mathcal{A}) +  \mathcal{B} \quad \Rightarrow \quad F = \dd f \wedge ( \dd t + \mathcal{A}) + f \mathcal{F} + \mathcal{H},  
\eea
where we have further defined $\mathcal{F} = \dd \mathcal{A}$ and $\mathcal{H} = \dd \mathcal{B}$. Up to the inclusion of the scalar $f$ and vector $\mathcal{B}$, this is the same ansatz as before. Now performing the dimensional reduction at the level of the action by simply inserting the ansatz, while leaving a demonstration of the consistency to appendix \ref{sec:KK}, we arrive at the 3D action: 
\bea
\label{red_act}
\mathcal{L} &=& \sqrt{\gamma} \biggl( R - \frac{1}{2 V^2} (\partial V)^2 + \frac{1}{4} V^2 \mathcal{F}^2 - \frac{1}{4} V (f \mathcal{F} + \mathcal{H})^2 + \frac{1}{2 V}  (\partial f)^2 \biggr).
\eea 
Varying this action with respect to the vectors $\mathcal{A}$ and $\mathcal{B}$, one gets the following equations of motion: 
\bea
\dd \left( V *_3 ( f \mathcal{F} + \mathcal{H}) \right) &=& 0, \\
\dd ( V^2 * \mathcal{F} ) - V \dd f \wedge * (f \mathcal{F} + \mathcal{H})  &=& 0.  
\eea
It is worth noting at this point that the exterior derivative of the second equation is consistent with the first equation and that truncating $ f= \mathcal{B} = 0$ we recover the equation (\ref{eq0}). With the equations of motion at hand, we are now in a position to dualise the vectors through the following redefinitions: 
\bea
V *_3 ( f \mathcal{F} + \mathcal{H}) &=& \dd \omega \\
V^2 *_3 \mathcal{F} &=& \dd \chi + \frac{1}{2} ( f \dd \omega - \omega \dd f), 
\eea
where the reader should note that the equations of motion still hold, so once again everything is consistent. Care should be taken with the sign of the kinetic terms since in 3D spacetime $*_3 *_3 = -1$, so we eventually end up with a scalar manifold of signature $(+, +, -, -)$. At this point we are in a position to import various existing older results in the literature \cite{Maison:1979kx, Kinnersley:1977pg, Kinnersley:1977ph, Galtsov:1995mb} (see \cite{Galtsov:2008zz} for an overview) through simple field redefinitions. Concretely, we can redefine as follows, 
\be
V = e^{\xi}, \quad f = \sqrt{2} v, \quad \omega = \sqrt{2} u
\ee
to bring the target spacetime manifold to the form \cite{Galtsov:1995mb}: 
\be
\label{galtsov_metric}
\dd s^2 = \frac{1}{2} [ \dd \xi^2 + e^{-2 \xi} ( \dd \chi + v \dd u - u \dd v )^2] - e^{-\xi} (  \dd v^2 +  \dd u^2). 
\ee
Once again, we recover the analysis of the previous section when $v = u = 0$. 

With the target spacetime identified, we can now enumerate the symmetries. The most trivial correspond to the three shift symmetries
\bea
\chi &\rightarrow& \chi + c, \nn
v &\rightarrow& v + c, \quad \chi \rightarrow \chi - c \, u, \nn
u &\rightarrow& u + c, \quad \chi \rightarrow \chi + c \, v, 
\eea
where $c$ is a constant. In addition, we have a pretty obvious rotational symmetry in the $(u, v)$ plane, which is the usual EM duality exchanging electric and magnetic fluxes in the Einstein-Maxwell theory. Although a little less obvious from the 4D perspective, we have the rescaling symmetry, 
\be
\xi \rightarrow \xi + c, \quad u \rightarrow e^{\frac{c}{2}} u, \quad v \rightarrow e^{\frac{c}{2}} v, \quad \chi \rightarrow e^{c} \chi. 
\ee
The remaining three transformations fleshing out the $SU(2,1)$ symmetry of the target spacetime are less obvious, but as explained in \cite{Galtsov:1995mb} are best described through the introduction of (complex) Ernst potentials: 
\be
\Phi = \frac{1}{\sqrt{2}} ( v + i u ), \quad \mathcal{E} = e^{\xi} + i \chi -\Phi \Phi^*.  
\ee
Then to read off the transformed target spacetime, one simply has to unravel a complex Harrison transformation   
\be
\Phi' = \frac{\Phi + \lambda \mathcal{E}}{1 - 2 \lambda^* \Phi - |\lambda|^2 \mathcal{E}}, \quad \mathcal{E}'  = \frac{\mathcal{E}}{1- 2 \lambda^* \Phi - |\lambda|^2 \mathcal{E}}, 
\ee
and a real Ehlers transformation, 
\be
\label{ehlers}
\Phi' = \frac{\Phi}{1 + i \gamma \mathcal{E}}, \quad \mathcal{E}' = \frac{\mathcal{E}}{1 + i \gamma \mathcal{E}}.  
\ee
into constituent components. 

\subsection{Maxwell fields in KS spacetimes}
 We begin our analysis by discussing the implications of a KS decomposition for the equations of motion of the Maxwell field strength. Note that the equations of motion following from the action (\ref{EM}) are 
\be
\dd * F = 0, \quad R^{\mu}_{ ~\nu} = \frac{1}{2} F^{\mu}_{~ \rho} F_{\nu}^{~\rho} - \frac{1}{8} \delta^{\mu}_{~\nu} F^2. 
\ee
First, let us observe that the Hodge dual in KS spacetimes may be explicitly written in terms of coordinates as 
\bea
(* F)_{\mu \nu} &=& \frac{1}{2!} \sqrt{g} \epsilon_{\mu \nu \rho \sigma} g^{\rho \lambda} g^{\sigma \epsilon} F_{\lambda \epsilon} \nn
&=&  \frac{1}{2!} \sqrt{\eta} \epsilon_{\mu \nu \rho \sigma} \left( F^{\rho \sigma}  - \phi k^{\sigma} k^{\epsilon} F^{\rho}_{~\epsilon} - \phi k^{\rho} k^{\lambda} F_{\lambda}^{~\sigma} \right), \nn
&=& \frac{1}{2!} \sqrt{\eta} \epsilon_{\mu \nu \rho \sigma} F^{\rho \sigma}, 
\eea
where in the second line we have used the fact the inverse of the metric (\ref{KS}) is linear in $\phi$ and $\textrm{det} (g) = \textrm{det} (\eta)$. We have also raised and lowered indices using the metric $\eta_{\mu \nu}$. In the third line we have used the result $k^{\mu} F_{\mu \nu} \propto k_{\nu}$ \footnote{One can show this by recalling $A = \phi k$ and using the null condition $k_{\mu} k^{\mu} = 0$ and the geodesic equation $k^{\mu} \partial_{\mu} k_{\nu} = 0$ to show that $k^{\mu} F_{\mu \nu} = (k^{\mu} \partial_{\mu} \phi) k_{\nu}$.} to convince ourselves that the $\phi$-dependent terms must cancel. Thus, the Hodge dual of a two-form on a KS spacetime is equivalent to the Hodge dual on flat spacetime. This means that the Maxwell field strength is a harmonic two-form on flat spacetime and this is a generic feature for all KS spacetimes. We expect similar conclusions to hold for differential forms of different dimensionality.

Next let us turn our focus to the r.h.s. of the Einstein equation. Using arguments similar to above, which essentially follow from the fact that $k$ is null, one can show that the r.h.s. reduces to the same expression but evaluated once again on flat spacetime, i. e. 
\be
R^{\mu}_{ ~\nu} = \left( \frac{1}{2} F^{\mu}_{~ \rho} F_{\nu}^{~\rho} - \frac{1}{8} \delta^{\mu}_{~\nu} F^2 \right) |_{g = \eta}. 
\ee
In summary, provided a KS description exists, the Maxwell equations of motion are reduced to those of flat spacetime. This constitutes a remarkable simplification.

\subsection{{A puzzle with the single copy}}
Having spelled out the implications for the Maxwell field, we turn our attention to what we will refer to as a generalised KS description  \cite{Vaidya:1947zz}, where we use ``generalised" in the sense that the scalar in the pure gravity KS ansatz picks up a component that depends on the Maxwell fields. Here, we quickly confirm that any black hole solution with purely diagonal metric components $g_{tt} = g^{rr} = f(r)$ can be brought to a KS form. We follow the treatment in \cite{Alawadhi:2019urr}. Consider the redefinition 
\be
\dd l = \dd t + \frac{\dd r}{f(r)}, 
\ee
so that the metric becomes 
\be
\dd s^2  = - f(r) \dd l^2 + 2 \dd l \dd r  + r^2 \dd s^2(S^2). 
\ee
Then further redefining $l = \bar{t} + r$, we can bring the metric to the generalised KS form: 
\be
\dd s^2 = - \dd \bar{t}^2 + \dd r^2 + r^2 \dd s^2(S^2) + \left( 1 - f(r) \right) ( \dd \bar{t} + \dd r)^2. 
\ee

Let us now be more specific and consider the Reissner-Nordstrom (RN) black hole where 
\be
f(r)  = 1 - \frac{2M}{r} + \frac{Q^2}{r^2}. 
\ee
As we show in appendix \ref{sec:dyon} this solution, as well as its dyonic generalisation, can be generated from Schwarzschild by employing a Harrison transformation.  In some sense, the Schwarzschild solution and the RN solution are no longer independent, they are related by a rotation in the effective 3D target manifold. 

From the perspective of scattering amplitudes this is an interesting result \footnote{We thank Donal O'Connell for highlighting this implication of the Harrison transformation.}. Let us explain why. In \cite{Moynihan:2019bor} it is shown that to recover the Schwarzschild and RN black hole solutions at leading order in $G$ and $e^2$ through minimal coupling one requires respectively tree-level and one-loop triangle leading singularities. In effect, the existence of a simple rotation from Schwarzschild to the RN solution suggests that some hint of the same symmetry should exist perturbatively at the level of scattering amplitudes. Thus, the existence of Harrison transformations also has implications for amplitudes.

Now this brings us to an interesting observation. As noted in \cite{Carrillo-Gonzalez:2017iyj}, we are in a position to split the scalar $\phi$ into a purely gravitational part $\phi_{g}$ and a part that is electromagnetic in origin $\phi_{em}$, $\phi = \phi_{g} + \phi_{em}$, where $\phi_g \sim r^{-1}$ and $\phi_{em} \sim r^{-2}$. As we have already shown the Maxwell field strength is a harmonic two-form on flat spacetime, which means that the r.h.s. of the Einstein equation scales as $F^2 \sim r^{-4}$. This term is cancelled by the $\phi_{em}$ term as we now explicitly demonstrate.

Recall that once a spacetime is written in KS format, the l.h.s of the Einstein equation becomes \cite{Monteiro:2014cda}: 
\bea
R^{0}_{~0} &=& \frac{1}{2} \nabla^2 \phi, \nn
R^{i}_{~0} &=& - \frac{1}{2} \partial_j \left[ \partial^i ( \phi k^j) - \partial^j (\phi k^i)\right], \nn
R^{i}_{~j} &=& \frac{1}{2} \partial_{l} \left[ \partial^i ( \phi k^l k_j) + \partial_j (\phi k^l k^i) - \partial^l (\phi k^i k_j) \right]. 
\eea
Explicitly for the RN solution and the first Einstein equation, we find the equation \footnote{In our conventions the field strength is $F = 2 \frac{Q}{r^2} \dd t \wedge \dd r$.}: 
\be
\frac{1}{2} \nabla^2 \phi_{g} - \frac{1}{2} \nabla^2 \left(  \frac{Q^2}{r^2} \right)  =  \frac{1}{2} F^{0}_{~\rho} F_{0}^{~\rho} - \frac{1}{8} F^2 = - \frac{(Q^2)}{r^4},  
\ee
where similar expressions exist for the remaining equations. Since the Laplace operator is defined with respect to flat spacetime, it is clear that the Einstein equation is satisfied through two different equations that are schematically of the form 
\be
\label{eoms}
\nabla^2 \phi_g \sim 0, \quad \nabla^2 \phi_{em} \sim \phi_{em}^2. 
\ee
While the first equation is just the expected term from the classical double copy for pure gravity \cite{Monteiro:2014cda}, where it is interpreted as the Maxwell equation, the second equation is intriguing. {Although we have focussed on the RN solution, essentially for illustrative purposes, it should be stressed that our observations here extend to dyonic black holes with spherical symmetry. Moreover, when the electric and magnetic charges agree, the corresponding solution is not only a solution to Einstein-Maxwell theory, but also Einstein-Maxwell-dilaton theory.}

Let us try to interpret these equations in terms of the putative single copy. To do so, we recall that at the perturbative level Einstein-Yang-Mills should be the double copy of Yang-Mills and {a bi-adjoint scalar field with a cubic potential} \cite{Chiodaroli:2014xia, Chiodaroli:2015rdg, Chiodaroli:2016jqw}. Let us start from the single copy action presented in equation (4.5) of \cite{Chiodaroli:2014xia}: 
\bea
\label{act_single}
\mathcal{L} &=& - \frac{1}{4} F^{\hat{a}}_{\mu \nu} F^{\mu \nu}_{\hat{a}} + \frac{1}{2} ( \DD_{\mu} \phi^a)^{\hat{a}} (\DD^{\mu} \phi^b)_{\hat{a}} \delta_{ab} + \frac{g g'}{3!} (i f_{\hat{a} \hat{b} \hat{c}}) F_{abc} \phi^{\hat{a} a} \phi^{\hat{b} b} \phi^{\hat{c} c}, 
\eea
where we have defined Yang-Mills field strengths and covariant derivatives
\be
F^{\hat{a}}_{\mu \nu} = \partial_{\mu} A^{\hat{a}}_{\nu} - \partial_{\nu} A^{\hat{a}}_{\mu}  + g f^{\hat{a}}_{ ~ \hat{b} \hat{c}}  A^{\hat{b}}_{\mu} A^{\hat{c}}_{\nu}, \quad (\DD_{\mu} \phi^a )^{\hat{a}} = \partial_{\mu} \phi^{\hat{a} a} + g f^{\hat{a}}_{~\hat{b} \hat{c}} A^{\hat{b}}_{\mu} \phi^{\hat{c} a}. 
\ee
 Following \cite{Chester:2017vcz}, we have dropped an $O(g^2)$ quartic scalar term in the action on the grounds that it is not needed for the double copy. One should note that $g'$ is arbitrary and dimensionful, while $g$ and $f^{\hat{a}}_{~ \hat{b} \hat{c}}$ denote the coupling and structure constants for the gauge group. The rank-three tensor $F_{abc}$ has entries that are given by the structure constants of a subgroup of $SO(n)$. We refer the reader to \cite{Chiodaroli:2014xia} for further details.

The corresponding equations of motion for the action (\ref{act_single}) may be expressed as 
\bea
\label{single_eq1} 0 &=& - \partial_{\mu}( \DD^{\mu} \phi_{a} )_{\hat{a}} + g f^{\hat{c}}_{~\hat{b} \hat{a}} A^{\hat{b}}_{\mu} ( \DD^{\mu} \phi_a)_{\hat{c}} + \frac{g g'}{2!} i f_{\hat{a} \hat{b} \hat{c}} F_{abc} \phi^{ \hat{b} b} \phi^{\hat{c} c}, \\
\label{single_eq2} 0 &=& \partial_{\mu} F_{\hat{a}}^{~\mu \nu} - g f^{\hat{c}}_{~\hat{b} \hat{a}} A^{\hat{b}}_{\mu} F_{\hat{c}}^{~\mu \nu} + g f^{\hat{b}}_{~\hat{a} \hat{c}} \phi^{\hat{c} a} (\DD^{\nu} \phi_{a})_{ \hat{b}}. 
\eea

{Let us now put these equations in context. First, one can consistently truncate out the scalars through $\phi^{\hat{a} a} = 0$ and one arrives at Yang-Mills theory through (\ref{single_eq2}), which is the expected single copy for pure gravity, at least in the classical double copy prescription \footnote{Strictly speaking, one would expect the bosonic sector of supergravity, i.e. $g_{\mu \nu}, B_{\mu \nu}$ and $\phi$, as in the perturbative double copy, but $B$ and $\phi$ {are not relevant in our example}.}. Alternatively, one could try to truncate out the gauge fields. Doing so, one recovers the biadjoint scalar equation from (\ref{single_eq1}), 
\be
\label{single_eq3}  \nabla^2 \phi_{\hat{a} a}  =  \frac{g g'}{2!} i f_{\hat{a} \hat{b} \hat{c}} F_{abc} \phi^{ \hat{b} b} \phi^{\hat{c} c}, 
\ee
but unfortunately the consistency of this truncation is spoiled by the final term in (\ref{single_eq2}) and one is left with a constraint on $\phi$, $f^{\hat{b}}_{~\hat{a} \hat{c}} \phi^{\hat{c} a} \partial^{\nu} \phi_{ \hat{b} a} = 0$. Modulo this constraint, one recovers the biadjoint scalar equation on the nose, which it should be stressed is the expected single copy for Yang-Mills theory. The key point we wish to stress here is that one can \textit{almost} disentangle the equations into the single copy for gravity and the single copy for Yang-Mills.}

{Now, let us recall that in the original classical double copy formulation \cite{Monteiro:2014cda}, one interprets the zeroth copy, or the $\nabla^2 \phi = 0$ equation, as simply a linearisation of the biadjoint scalar equation (\ref{single_eq3}). It is valid to ask if one can go further. Here, we can follow the analysis of \cite{White:2016jzc} and adopt a spherical ansatz for the scalar $\phi^{\hat{a} a} = \delta^{\hat{a} a} \phi$, while at the same time identifying the structure constants $f_{\hat{a} \hat{b} \hat{c}}$ and $i F_{abc}$, which for simplicity could be identified with the structure constants of the Lie algebra $\frak{g} = \frak{su}(2)$ \footnote{As noted in \cite{White:2016jzc}, the more general choice $\phi^{\hat{a} a} = \chi^{\hat{a}} \xi^{a}$ leads to a vanishing of the cubic term and a resulting free field theory.}. This condition is enough to solve the constraint remaining from (\ref{single_eq2}), so that (\ref{single_eq1}) reduces to the required additional equation on the right hand side of (\ref{eoms}).}

{Note, throughout the above discussion we omitted the gauge fields, but these can be reintroduced and truncated through the choice $A^{a} = A \Rightarrow F^{a} = F$. This would seem to interfere with the biadjoint scalar equation (\ref{single_eq1}), but for the choice of configurations we consider, namely spherically symmetric charged black holes in Einstein-Maxwell theory, one can confirm that $\partial_{\mu} A^{\mu} = A^{\mu} \partial_{\mu} \phi = 0$ is always true. To appreciate this fact, note that $\phi$ is a function of the radial direction, whereas $A$ is the Maxwell field inferred from the metric, which for any charged black hole related to Schwarzschild through a Harrison-Ehlers transformation will only possess $A_t$ and $A_{\phi}$ components.}

{Let us summarise our discussions here. We observed that any spherically symmetric solution to Einstein-Maxwell theory could be put generalised KS form. Doing so, one observes that the mass and the charges have to cancel separately through two distinct equations. One of these is the original equation from the classical double copy and this suffices to explain metrics with $r^{-1}$ scaling. We have argued that the additional equation can be recovered from the equations of motion of the putative single copy to Einstein-Yang-Mills via a well-defined truncation procedure. In particular, this truncation allows one to avoid the usual linearisation of the biadjoint scalar equation \cite{Monteiro:2014cda}.}

\subsection{Ehlers transformation} 
\label{sec:ehlersEM}
Our goal in this subsection is to show how the nontrivial Ehlers transformation (\ref{ehlers}) is EM duality in the extended setting of Einstein-Maxwell theory. This should be contrasted with the trivial form of EM duality where one exploits the rotational symmetry in the $(v, u)$ directions of the target spacetime.  Concretely,   let us start with the dyonic solution (\ref{dyon})  and generate a new geometry with electric, magnetic and NUT charge. After performing an Ehlers transformation (\ref{ehlers}) and a rescaling, the final data becomes: 
\bea
\e^{\xi} &=& \frac{(r^2 - 2 M r + Q^2 + P^2) e^{c}}{r^2 + \gamma^2 (r-2 M)^2}, \quad \chi = - \frac{\gamma (r-2M)^2 e^{c}}{r^2 + \gamma^2 (r-2 M)^2}, \nn
v &=& - \frac{\sqrt{2} ( Q r + \gamma P (r-2M) ) e^{\frac{c}{2}}}{r^2 + \gamma^2 (r-2 M)^2}, \quad u = - \frac{\sqrt{2} ( P r -\gamma Q (r-2M) ) e^{\frac{c}{2}}}{r^2 + \gamma^2 (r-2 M)^2}. 
\eea
Translating from the 3D target spacetime to the 4D solution to Einstein-Maxwell, we arrive at the following solution: 
\bea
\dd s^2 &=& - e^{\xi} (\dd t + \mathcal{A})^2 + e^{-\xi} \left( \dd r^2 + r^2 f(r) \dd s^2(S^2) \right), \nn
A &=& \sqrt{2} v ( \dd t + \mathcal{A} ) + \mathcal{B} 
\eea
where we have defined
\bea
\mathcal{A} = 4 M e^{-c} \gamma \cos \theta \dd \phi, \quad \mathcal{B} = - 2 e^{- \frac{c}{2}} ( P - \gamma Q) \cos \theta \dd \phi. 
\eea
By performing the following transformations \cite{Alawadhi:2019urr}, 
\be
e^{c} = 1 + \gamma^2, \quad N = \frac{2 \gamma M}{1 + \gamma^2}, \quad r - N \gamma = \rho, \quad  M' = M \frac{1-\gamma^2}{1+\gamma^2}. 
\ee
we can bring it to the form:
\bea
\dd s^2 &=& - \frac{\rho^2 - 2 M' \rho -N^2 + Q^2 + P^2}{\rho^2 + N^2} (\dd t + 2 N \cos \theta \dd \phi)^2  \nn &+& \frac{\rho^2 + N^2} {\rho^2 - 2 M' \rho -N^2 + Q^2 + P^2}\dd \rho^2 +  ( \rho^2 + N^2) \dd s^2(S^2), \\
A &=&  - 2 \left [  \frac{\rho (Q+ \gamma P) - N (P - \gamma Q)}{\rho^2 + N^2} ( \dd t + 2 N \cos \theta \dd \phi) + ( P - \gamma Q) \cos \theta \dd \phi \right], \nonumber
\eea
where we have performed the following further redefinitions 
\be
\frac{Q + \gamma P}{\sqrt{1+\gamma^2} } \rightarrow Q, \quad \frac{P - \gamma Q}{\sqrt{1+\gamma^2} } \rightarrow P, 
\ee
so that we recover the RN-Taub-NUT solution of \cite{AlonsoAlberca:2000cs} with $g=0$ (see also \cite{Mann:2005mb} for the purely electric solution).

At this point we are again in a position to comment on the Maxwell charges in the double copy. However, there are some small differences, which we now outline, otherwise the basic idea is the same. In our earlier section we showed that (\ref{Fm}) held in the absence of a Maxwell field. As explained above, the Maxwell field in the double copy formalism is essentially a Maxwell field inferred from the metric, which sources a scalar potential $\phi_g$ in a generalised KS description. Since this is purely a quantity we define at the level of the metric, the Maxwell fields in Einstein-Maxwell do not affect this definition. For this reason, (\ref{Fm}) is generalised to 
\be
\label{Fmag_EM}
*_3 F_{\textrm{mag}} =  V^2 *_3 \mathcal{F} = \dd \chi + v \dd u - u \dd v, 
\ee
where as before Hodge duality is performed on different spaces. It is worth noting that in the original dyonic solution (\ref{dyon}) this term is not sourced, so the Maxwell field in the double copy is purely electric. However, once we perform an Ehlers transformation, as we have seen above, $\chi, u$ and $v$ become non-zero, so this term makes a contribution.  Taking into account the rescaling, we find that the electric and magnetic flux are 
\be
Q_e =  \frac{1}{4 \pi} \int_{S^2} *_3 \dd V =  2 M', \quad Q_m = \frac{1}{4 \pi} \int_{S^2} *_3 ( \dd \chi + v \dd u -u \dd v) = - 2 N,
\ee
where it can checked that $u, v$ have the correct asymptotic form, i. e. $u, v \rightarrow 0$ as $r \rightarrow \infty$, thus ensuring that our defintion of the charges is once again consistent with the scalar equations of motion. Note, these charges are purely gravitational and should not be confused with $P, Q$, which are electromagnetic in nature.

It is worth noting that since $N^2 + M'^2 = M^2$, or alternatively since the NUT charge $N$ combines with the new mass $M'$ to recover the original mass of the black hole $M$, both the electric and magnetic Maxwell charges have been transformed from the original Schwarzschild geometry. Finally, following the arguments similar to \cite{Chong:2004hw}, where a class of solutions to Einstein-Maxwell in Plebanski formalism \cite{Plebanski:1975xfb} are studied, one can convince oneself that the above metric can be brought to a generalised double KS description.

\section{Conclusions} 
In the earlier part of this work we married the KS ansatz of pure gravity with a natural 4D to 3D dimensional reduction, which allowed us to identify the Maxwell field strengths of the double copy formalism directly in terms of the scalars parametrising a hyperbolic coset geometry in 3D. As we have argued, this can be done for generic spacetime geometries and it is the rotation of the scalars under a $U(1) \subset SL(2, \mathbb{R})$ that is mapped to EM duality in the Maxwell fluxes of the double copy formalism. We believe our work clarifies and generalises to stationary spacetimes admitting a (double) KS form, the findings presented earlier in \cite{Alawadhi:2019urr}. 
 
In the latter part of this work, we extended our findings to Einstein-Maxwell theory in 4D. To do so, we identified the 3D $\sigma$-model with $SU(2,1)$ symmetry, and {tried} to interpret the equations of motion of Einstein-Maxwell theory, {at least in the context of spherically symmetric charged black holes}, in terms of the equations of motion of the putative single copy. {In particular, starting form the putative single copy for Einstein-Yang-Mills \cite{Chiodaroli:2014xia, Chester:2017vcz}, we showed that a truncation exists whereby one not only recovers the usual classical double copy prescription that allows for $r^{-1}$ terms in the metric, but also an additional equation that corresponds to the $r^{-2}$ terms. As we explained, the latter is a truncation of the biadjoint scalar equation constituting the expected single copy for Yang-Mills theory, thus generalising the zeroth copy of ref. \cite{Monteiro:2014cda}. That being said, it should be stressed here that we are not claiming that Einstein-Maxwell has a single copy description, only that one can find an interpretation of the the equations of motion for spherically symmetric charged black holes in Einstein-Maxwell within the putative single copy for Einstein-Yang-Mills theory. This is simply a statement about the existence of a truncation, but whether or not it is coincidental requires further investigation. } 

Our work raises a number of interesting future directions. First, it is clear that EM duality in the double copy can be realised in terms of BMS symmetries \cite{Huang:2019cja} and our analysis shows EM duality is also related to Ehlers transformations (see also \cite{Alawadhi:2019urr}), so the task remains to connect Ehlers transformations to BMS symmetries directly in pure gravity, before potentially extending to Einstein-Maxwell theory or equivalent. {On that note, an interesting recent paper \cite{Emond:2020lwi} discusses dualities in linearised gravity and it is conceivable that this is the Ehlers transformation at the non-linear level once a Killing direction is assumed.} Moreover, the connection in section 3.2  is intriguing and we should endeavour to {extend it to Einstein-Maxwell-dilaton theory, a setting where the (classical) double copy is expected to be on firmer footing.} 

It would be interesting to extend the results presented here to asymptotically (anti)-de Sitter spacetimes. That being said, the simplest generalisation of introducing a cosmological constant does not work. To see why, let us return to (\ref{action}) and observe that the inclusion of a cosmological constant in 4D leads to a 3D cosmological constant dressed by a $V^{-1}$ factor. As observed in \cite{Petkou:2015fvh} (see also \cite{Astorino:2012zm}), demanding that the 3D action is invariant forces one to consider $SL(2, \mathbb{R})$ transformations that are pure gauge (\ref{puregauge}). Nevertheless, extensions to different dimensions appear pretty immediate. Indeed, there are numerous examples of supergravity theories - where the classical double copy story is being actively studied \cite{Lee:2018gxc, Cho:2019ype, Kim:2019jwm} - that can be truncated to scalar sectors and similar symmetries arise.

\section*{Acknowledgement}
We thank Ilya Bakhmatov, David Berman, Kanghoon Lee, Nathan Moynihan, Miok Park and Shahin Sheikh-Jabbari for correspondence and discussion. We thank Donal O'Connell for sharing his expertise on the double copy formalism. E. \'O C thanks Yasha Neiman and OIST for hospitality during the write-up process. This work was supported in part by the Korea Ministry of Science, ICT \& Future Planning, Gyeongsangbuk-do and Pohang City and  the National Natural Science
Foundation of China, Project 11675244.

\appendix

\section{Symmetries the coset $SL(2, \mathbb{R})/U(1)$}
\label{sec:coset}
In this section, we illustrate how the symmetries are manifest at the level of the coset $SL(2, \mathbb{R})/U(1)$. Consider the matrices 
\be
t_0 = \left( \begin{array}{cc} 1 & 0 \\ 0 & - 1 \end{array} \right), \quad t_{+} =   \left( \begin{array}{cc} 0 & 1 \\ 0 & 0 \end{array} \right), \quad t_{-} =  \left( \begin{array}{cc} 0 & 0 \\ 1 &  0 \end{array} \right), 
\ee
which generate the Lie algebra $\frak{sl}(2)$: 
\be
[ t_0, t_+] = 2 t_+, \quad [t_0, t_-] = - 2 t_{-}, \quad [t_+, t_-] = t_0.  
\ee 
Exponentiating these matrices, we get elements of the Lie group $SL(2, \mathbb{R})$, 
\be
e^{\alpha t_0} = \left( \begin{array}{cc} e^{\alpha} & 0 \\ 0 &  e^{-\alpha} \end{array} \right), \quad e^{\beta t_+} = \left( \begin{array}{cc} 1 & \beta \\ 0 & 1 \end{array} \right), \quad e^{\gamma t_-} = \left( \begin{array}{cc} 1 & 0 \\ \gamma & 1 \end{array} \right). 
\ee
Note, these are clearly all of the form 
\be
\label{element_sl2}
\left( \begin{array}{cc} a & b \\ c & d \end{array} \right) \in SL(2,R), \quad ad - bc = 1.
\ee
From the matrix, 
\be
\mathcal{V} = e^{\chi t_+} e^{ \ln \sqrt{V} t_0} = \left( \begin{array}{cc} V^{\frac{1}{2}} & V^{-\frac{1}{2}} \chi  \\ 0 & V^{-\frac{1}{2}}  \end{array} \right), 
\ee
we can define the current $ J = \mathcal{V}^{-1} \dd \mathcal{V}$, from where we can further define the target space metric through $ \dd s^2 = \Tr ( P^2)$, where $P$ denotes the symmetric part of $J$, $P \equiv \frac{1}{2} ( J + J^T)$. The finite transformations come from 
\be
\mathcal{M}' = g \mathcal{M} g^{T}, 
\ee
where $g \in SL(2, \mathbb{R})$ and we have defined $\mathcal{M} = \mathcal{V} \mathcal{V}^{T}$. Concretely, we have 
\be
\left( \begin{array}{cc} V' + \frac{\chi'^2}{V'} & \frac{\chi'}{V'}  \\ \frac{\chi'}{V'} & \frac{1}{V'} \end{array} \right) = \left( \begin{array}{cc} a & b \\ c & d \end{array} \right) \left( \begin{array}{cc} V + \frac{\chi^2}{V} & \frac{\chi}{V}  \\ \frac{\chi}{V} & \frac{1}{V} \end{array} \right) \left( \begin{array}{cc} a & c \\ b & d \end{array} \right). 
\ee
One can check that this is equivalent to 
\be
\tau' = \frac{a \tau + b}{c \tau + d}, \quad \tau \equiv \chi + i V. 
\ee
This provides a realisation of the symmetries of the hyperbolic space $H^2$ starting from the coset description. Here, it is clear that $t_0$ is generating scale transformations, $t_+$ corresponds to pure gauge transformations, while it is $t_-$ that is generating the non-trivial Ehlers transformations.

\section{General $SL(2,\mathbb{R})$ transformation} 
\label{sec:gensl2}
In this section we comment on the general $SL(2, \mathbb{R})$ transformation applied to Schwarzschild with a goal to convince ourselves that of the three unconstrained $SL(2, \mathbb{R})$ parameters, only one is relevant after various redefinitions. Recall the most general form of a $SL(2,\mathbb{R})$ transformation is given by \eqref{sl2R}. For the Schwarzschild solution we have
\be
V=1-\frac{2M}{r}, \quad \chi=0, 
\ee
so under the  $SL(2,\mathbb{R})$ transformation we get following expressions for $V'$ and $\chi'$,
\be
\label{gensol}
V'=\frac{r(r-2M)}{c^2(r-2M)^2+d^2 r^2},\quad \chi'=\frac{ac(r-2M)^2+bd r^2}{c^2(r-2M)^2+d^2 r^2}\;.
\ee
Using \eqref{key} the two form $\mathcal{F}$ takes following form
\be
\quad \mathcal{F}= 4dcM\vol (S^2)\;.
\ee
Finally the 4D metric can be written as
\be\label{4dm}
\dd s^2=-\frac{r(r-2M)}{c^2(r-2M)^2+d^2r^2}\left(\dd t + 4Mdc\cos\theta \dd \varphi\right)^2+\frac{c^2(r-2M)^2+d^2r^2}{r(r-2M)} \dd s_3^2
\ee
where we have defined, 
\be
\dd s_3^2=\dd r^2+r(r-2M) ( \dd \theta^2+ \sin^2\theta \dd \varphi^2). 
\ee
In above metric only two of independent parameters of $SL(2,\mathbb{R})$ appear. Indeed one can show that one of these parameters can be also eliminated by a shift and a rescaling of both the radial and time coordinates. If we define two positive parameters $r_\pm$ by
\be
r_+=\frac{2Mc^2}{\sqrt{c^2+d^2}},\quad r_-=\frac{2Md^2}{\sqrt{c^2+d^2}}.
\ee
After a shift and scaling 
\be
r\rightarrow \frac{r+r_+}{\sqrt{c^2+d^2}},\qquad t\rightarrow \frac{t}{\sqrt{c^2+d^2}}\ee
metric \eqref{4dm}  take following form
\be
\dd s^2=-f(r)\left(\dd t -2\sqrt{r_+r_-} \cos\theta \dd \varphi \right)^2+\frac{\dd r^2}{f(r)}+(r^2+r_+r_-)\dd s^2 (S^2), 
\ee
where function $f$ is defined by
\be
f(r)=\frac{(r+r_+)(r-r_-)}{r^2+r_+r_-}
\ee
This is the metric of the Taub-NUT space time. 
  
\section{Double KS} 
\label{sec:doubleKS}
In this section we provide some details to support the identity (\ref{Fm}). We work with a double KS ansatz for greater generality. Consider the double KS ansatz: 
\be
g_{\mu \nu} = \eta_{\mu \nu} + \phi k_{\mu} k_{\nu} + \psi l_{\mu} l_{\nu}, 
\ee
where $k_\mu$ and $l_\mu$ satisfy following equations,
\bea\label{const}
&& \eta^{\mu\nu}k_\mu k_\nu=g^{\mu\nu}k_\mu k_\nu =0,\quad \eta^{\mu\nu}l_\mu l_\nu=g^{\mu\nu}l_\mu l_\nu =0, \nonumber \\
&&  \eta^{\mu\nu}k_\mu l_\nu= g^{\mu\nu}k_\mu l_\nu=0,\quad k_\mu \partial^\mu k_\nu =0,\quad l_\mu \partial^\mu l_\nu =0 \,
\eea
Assuming $\partial_t$ is a Killing direction, we can always write 
\be
k = \dd t + \tilde{k}, \quad l = \dd t + \tilde{l}. 
\ee
Rewriting everything in terms of the earlier ansatz (\ref{ehlers}) gives
\be\label{A}
V=1-\phi-\psi, \qquad A=-V^{-1}\left( \phi \tilde k +\psi \tilde l \right)
\ee
and the 3D metric $\dd s_3^2$ becomes
\be
\dd s_3^2=\gamma_{mn}\dd x^m \dd x^n=V (\dd x_i^2 +\phi \tilde k^2 +\psi \tilde l^2)   +  ( \phi \tilde k +\psi \tilde l )^2
\ee
To find the Hodge dual, first we need to invert the above metric. It is easy to show that the inverse metric is
\be
\gamma^{mn} = (1- \phi - \psi)^{-1} \left( \delta^{mn} - \phi \tilde{k}^m \tilde{k}^n - \psi \tilde{l}^m \tilde{l}^n \right). 
\ee
where we defined $\tilde k^m$ and $\tilde l^m$ by
\be
\tilde k^m =  \delta^{mn} \tilde k_n,\quad \tilde l^m =  \delta^{mn}  \tilde l_n. 
\ee

Now, using \eqref{A} and \eqref{const} we get
\be
(V^2 *_3 \mathcal{F})_p=-2\partial^m(\phi  \tilde{k}^n+ \psi  \tilde{l}^n)\epsilon_{mnp}
\ee
which can be further rewritten as (\ref{Fm}). Although we have not performed the calculation, there is nothing that suggests the same analysis will not work for a KS ansatz with three null vectors. 

\section{Consistency of dimensional reduction}
\label{sec:KK} 
In this section, we show that the dimensional reduction of 4D Einstein-Maxwell theory on a temporal direction leads to the four-dimensional target spacetime in the text. We will perform this reduction at the level of the action and equations of motion (EOMs), thereby demonstrating consistency. 
The EOMs of the action (\ref{EM}) are 
\be
R_{\mu \nu}  - \frac{1}{2} \left( F_{\mu \rho} F_{\nu}^{~\rho} - \frac{1}{4} g_{\mu \nu} F^2 \right)  = \dd *_4 F = 0. 
\ee

Now, we can reduce to 3D through the ansatz \footnote{This is the same as the ansatz in the text up to the replacement $e^{2V} \rightarrow V$.}: 
\bea
\dd s^2_4 &=& - e^{2 V} ( \dd t + \mathcal{A})^2 + \dd s^2_3, \nn
A &=& f ( \dd t + \mathcal{A}) +  \mathcal{B} \quad \Rightarrow \quad F = \dd f \wedge ( \dd t + \mathcal{A}) + f \mathcal{F} + \mathcal{H},  
\eea
where we have further defined $\mathcal{F} = \dd \mathcal{A}$ and $\mathcal{H} = \dd \mathcal{B}$. 

For the reduction to 3D, we have the Ricci tensor, 
\bea
R_{\alpha \beta} &=& \bar{R}_{\alpha \beta} -  \nabla_{\beta} \nabla_{\alpha} V - \partial_{\alpha} V \partial_{\beta} V + \frac{1}{2} e^{2 V} \mathcal{F}_{\alpha \gamma} \mathcal{F}_{\beta}^{~\gamma} \nn
R_{\alpha 0} &=&  \frac{1}{2} e^{-2 V} \nabla_{\gamma} ( e^{3 V} \mathcal{F}^{\gamma}_{~ \alpha} ), \nn
R_{00} &=& \nabla_{\gamma} \nabla^{\gamma} V +  \partial_{\gamma} V \partial^{\gamma} V + \frac{1}{4} e^{2 V} \mathcal{F}_{\alpha \beta} \mathcal{F}^{\alpha \beta}. 
\eea

Doing the reduction directly at the level of the action (\ref{EM}), we get: 
\bea
\label{act}
\mathcal{L}_3 = \sqrt{g_3} e^{V} \biggl( \bar{R} + \frac{1}{4} e^{2 V} \mathcal{F}^2 - \frac{1}{4} (f \mathcal{F} + \mathcal{H})^2 + \frac{1}{2} e^{ -2 V} (\partial f)^2 \biggr). 
\eea
Note that this action is not in Einstein frame, but performing the conformal transformation $g_3 \rightarrow e^{-2 V} \gamma$ brings it to the form quoted in the text once $V$ is redefined. It turns out that the truncation is consistent as we now demonstrate. 

First, observe that 4D Maxwell EOM leads to the two equations in 3D: 
\bea
\label{eq1} \dd \left( e^{V} *_3 ( f \mathcal{F} + \mathcal{H}) \right) &=& 0, \\
\label{eq2} \dd \left( e^{ -V} *_3 \dd f \right) + e^{V} *_3 (f \mathcal{F} + \mathcal{H}) \wedge \mathcal{F} &=& 0.  
\eea
These equations follow from the action (\ref{act}) upon varying with respect to $\mathcal{B}$ and $f$, respectively, so the action passes the first test. Note, the first equation will allow us introduce a scalar.

The remaining Einstein equation becomes: 
\bea
\label{eq3} \nabla^2 V +  (\partial V)^2 + \frac{1}{4} e^{2 V} \mathcal{F}^2 - \frac{1}{4} e^{- 2V} (\partial f)^2 - \frac{1}{8} (f \mathcal{F}+\mathcal{H})^2 &=& 0, \\
\label{eq4} \frac{1}{2} e^{-2 V} \nabla_{\gamma} ( e^{3 V} \mathcal{F}^{\gamma}_{~\alpha} ) + \frac{1}{2} e^{ - V} \partial_{\gamma} f ( f \tilde{F} + H)_{\alpha}^{~\gamma} &=& 0, \\
\label{eq5} \bar{R}_{\alpha \beta} -  \nabla_{\beta} \nabla_{\alpha} V - \partial_{\alpha} V \partial_{\beta} V  + \frac{1}{2} e^{-2V} \partial_{\alpha} f \partial_{\beta} f  + \frac{1}{2} e^{2 V} \mathcal{F}_{\alpha \gamma} \mathcal{F}_{\beta}^{~\gamma} \nn - \frac{1}{2} (f \mathcal{F} + \mathcal{H})_{\alpha \gamma} (f \mathcal{F} + \mathcal{H})_{\alpha}^{~ \gamma}   &=& 0. 
\eea
One can check that (\ref{eq3}) follows from the action (\ref{act}) by varying $V$, while (\ref{eq4}) follows from (\ref{act}) by varying with respect to $\mathcal{A}$. Finally, it can be checked that the Einstein equation follows from the action. This demonstrates the consistency of the reduction, which is modulo a conformal transformation the same as the action (\ref{red_act}) quoted in the text. 

\section{Harrison transformation}
\label{sec:dyon}
Since these transformations may look new to a hep-th readership, let's get oriented by describing an explicit example. To begin, let us first assume $\xi \neq 0$, which is enough to describe a Schwarzschild solution, and generate a dyonic solution through a complex Harrison transformation. To preserve asymptotic flatness, as we did previously for the Ehlers transformation in pure gravity, we will also perform a scale transformation. More concretely, we will consider 
\be
\lambda = \kappa e^{i \alpha}. 
\ee

After the transformation, the solution is
\be
v = \frac{\kappa \cos \alpha \sqrt{2} e^{\xi_0} e^{\frac{c}{2}}}{1 - \kappa^2 e^{\xi_0}}, \quad u = \frac{\kappa \sin \alpha \sqrt{2} e^{\xi_0} e^{\frac{c}{2}}}{1 - \kappa^2 e^{\xi_0}},, \quad e^{\xi} = \frac{e^{\xi_0} e^{c}}{(1-\kappa^2 e^{\xi_0})^2}, 
\ee
where $c$ is a scaling parameter we have introduced and $\xi_0$ denotes the original data, in this case the Schwarzschild solution: 
\be
e^{\xi_0} = 1- \frac{2 M}{r}. 
\ee
The first thing to note is that this transformation is a symmetry of the target spacetime and one can check that 
\be
\frac{1}{2} \dd \xi^2 - e^{-\xi} ( \dd v^2 + \dd u^2) =  \frac{1}{2} \dd \xi_0^2.  
\ee
Since the 3D effective action is invariant, this guarantees a new solution in 4D and the task remains to identify the explicit form of the final solution. Evaluating all expressions, and performing Hodge dualities where necessary to identify the vector fields, we find that the final solution may be expressed as, 
\bea
\dd s^2 &=& - \frac{e^{c} r(r-2 M )}{(r(1-\kappa^2) + 2 M \kappa^2)^2} \dd t^2 + \frac{(r(1-\kappa^2) + 2 M \kappa^2)^2}{e^{c} r(r-2M)} \left[  \dd r^2 + r(r-2M) \dd s^2(S^2) \right], \nn
A &=& \frac{2 e^{\frac{c}{2}} \kappa \cos \alpha (r-2 M)}{r(1-\kappa^2) + 2 M \kappa^2} \dd t - e^{-\frac{c}{2}} \kappa \sin \alpha \, 4 M \cos \theta \dd \phi. 
\eea
This is still rather unsightly and to bring it to a more appealing form, one should consider the following change in the radial coordinate and accompanying judicious choice for the rescaling, 
\be
e^{\frac{c}{2}} \tilde{r} = r(1-\kappa^2) + 2 M \kappa^2, \quad e^{c} = (1- \kappa^2)^2. 
\ee
With these substitutions, we can simply drop tildes on the radial coordinate and recast the solution as  
\bea
\dd s^2 &=& - f(r) \dd t^2 + \frac{1}{f(r)} \dd r^2 + r^2 \dd s^2(S^2), \nn
A &=& 2 \kappa \cos \alpha \left( 1- \frac{2 M}{r (1- \kappa^2)} \right) \dd t -  \frac{\kappa}{(1-\kappa^2)} \sin \alpha \, 4 M \cos \theta \dd \phi, 
\eea
where we have further defined 
\be
f(r) = 1 - \frac{ 2M (1+\kappa^2)}{r (1- \kappa^2)} + \frac{4 M^2 \kappa^2}{r^2 (1-\kappa^2)^2}. 
\ee
Observe that the rescaling was instrumental to recover flat asymptotics and that the final solution can be brought to a more familiar form through the redefinitions, 
\be
M' = \frac{M (1+\kappa^2)}{(1-\kappa^2)}, \quad Q = \frac{\kappa \cos \alpha 2 M}{(1-\kappa^2)}, \quad P = \frac{\kappa \sin \alpha 2 M}{(1-\kappa^2)}, 
\ee
so that we arrive at the final expression: 
\bea
\label{dyon}
\dd s^2 &=& - f(r) \dd t^2  + \frac{1}{f(r)} \dd r^2 + r^2 \dd s^2(S^2), \nn
A &=& - \frac{2 Q}{r} \dd t - 2 P \cos \theta \dd \phi, \quad f(r) =  1 - \frac{2 M}{r} + \frac{Q^2 + P^2}{r^2}. 
\eea
Note, we have dropped a prime on $M$ and also a constant in the electric component of $A$ that was pure gauge. 

On the whole, this is more or less as may have been expected. We have seen that when pure Einstein gravity is coupled to Maxwell theory in 4D, we have a larger class of hidden symmetries upon dimensional reduction on a $U(1)$ Killing direction. Within this class, one finds the class of transformations originally identified by Harrison \cite{Harrison}, which provides a means to generate charged black hole solutions from the Schwarzschild solution. Here, we have opted for a complex transformation, so that the resulting geometry is dyonic, but a real Harrison transformation in tandem with a rotation in the $(u, v)$-plane of the target spacetime achieves the same result. We emphasise once again the role of a rescaling transformation in maintaining the asymptotics.


\begin{thebibliography}{99}


\bibitem{Monteiro:2014cda}
R.~Monteiro, D.~O'Connell, and C.~D. White, ``{Black holes and the double
  copy},'' {\em JHEP} {\bf 1412} (2014) 056,
\href{http://www.arXiv.org/abs/1410.0239}{{ 1410.0239}}.


\bibitem{Bern:2008qj} 
  Z.~Bern, J.~J.~M.~Carrasco and H.~Johansson,
  ``New Relations for Gauge-Theory Amplitudes,''
  Phys.\ Rev.\ D {\bf 78}, 085011 (2008)
  [arXiv:0805.3993 [hep-ph]].

\bibitem{Bern:2010ue} 
  Z.~Bern, J.~J.~M.~Carrasco and H.~Johansson,
  ``Perturbative Quantum Gravity as a Double Copy of Gauge Theory,''
  Phys.\ Rev.\ Lett.\  {\bf 105}, 061602 (2010)
  [arXiv:1004.0476 [hep-th]].

\bibitem{Bern:2010yg} 
  Z.~Bern, T.~Dennen, Y.~t.~Huang and M.~Kiermaier,
  ``Gravity as the Square of Gauge Theory,''
  Phys.\ Rev.\ D {\bf 82}, 065003 (2010)
  [arXiv:1004.0693 [hep-th]].

\bibitem{Bern:2019prr} 
  Z.~Bern, J.~J.~Carrasco, M.~Chiodaroli, H.~Johansson and R.~Roiban,
  arXiv:1909.01358 [hep-th].
  
  \bibitem{Sabharwal:2019ngs}
S.~Sabharwal and J.~W. Dalhuisen, ``{Anti-Self-Dual Spacetimes, Gravitational
  Instantons and Knotted Zeros of the Weyl Tensor},'' {\em JHEP} {\bf 07}
  (2019) 004,
\href{http://www.arXiv.org/abs/1904.06030}{{ 1904.06030}}.

\bibitem{Anastasiou:2014qba}
A.~Anastasiou, L.~Borsten, M.~J. Duff, L.~J. Hughes, and S.~Nagy, ``{Yang-Mills
  origin of gravitational symmetries},'' {\em Phys. Rev. Lett.} {\bf 113}
  (2014), no.~23, 231606,
\href{http://www.arXiv.org/abs/1408.4434}{{ 1408.4434}}.

\bibitem{Borsten:2015pla}
L.~Borsten and M.~J. Duff, ``{Gravity as the square of Yang-Mills?},'' {\em
  Phys. Scripta} {\bf 90} (2015) 108012,
\href{http://www.arXiv.org/abs/1602.08267}{{ 1602.08267}}.

\bibitem{Ridgway:2015fdl} 
  A.~K.~Ridgway and M.~B.~Wise,
  ``Static Spherically Symmetric Kerr-Schild Metrics and Implications for the Classical Double Copy,''
  Phys.\ Rev.\ D {\bf 94}, no. 4, 044023 (2016)
  [arXiv:1512.02243 [hep-th]].

\bibitem{Luna:2016due} 
  A.~Luna, R.~Monteiro, I.~Nicholson, D.~O'Connell and C.~D.~White,
  ``The double copy: Bremsstrahlung and accelerating black holes,''
  JHEP {\bf 1606}, 023 (2016)
  [arXiv:1603.05737 [hep-th]].
  
\bibitem{Carrillo-Gonzalez:2017iyj} 
  M.~Carrillo-Gonzalez, R.~Penco and M.~Trodden,
  ``The classical double copy in maximally symmetric spacetimes,''
  JHEP {\bf 1804}, 028 (2018)
  [arXiv:1711.01296 [hep-th]].

\bibitem{Anastasiou:2016csv}
A.~Anastasiou, L.~Borsten, M.~J. Duff, M.~J. Hughes, A.~Marrani, S.~Nagy, and
  M.~Zoccali, ``{Twin supergravities from Yang-Mills theory squared},'' {\em
  Phys. Rev.} {\bf D96} (2017), no.~2, 026013,
\href{http://www.arXiv.org/abs/1610.07192}{{ 1610.07192}}.

\bibitem{Anastasiou:2017nsz}
A.~Anastasiou, L.~Borsten, M.~J. Duff, A.~Marrani, S.~Nagy, and M.~Zoccali,
  ``{Are all supergravity theories Yang-Mills squared?},''
\href{http://www.arXiv.org/abs/1707.03234}{{ 1707.03234}}.

\bibitem{Cardoso:2016ngt}
G.~L. Cardoso, S.~Nagy, and S.~Nampuri, ``{A double copy for $ \mathcal{N}=2 $
  supergravity: a linearised tale told on-shell},'' {\em JHEP} {\bf 10} (2016)
  127,
\href{http://www.arXiv.org/abs/1609.05022}{{ 1609.05022}}.

\bibitem{Cardoso:2016amd} 
  G.~Cardoso, S.~Nagy and S.~Nampuri,
  ``Multi-centered $ \mathcal{N}=2 $ BPS black holes: a double copy description,''
  JHEP {\bf 1704}, 037 (2017)
  [arXiv:1611.04409 [hep-th]].

\bibitem{Borsten:2017jpt}
L.~Borsten, ``{On $D=6$, $\mathcal{N}=(2,0)$ and $\mathcal{N}=(4,0)$
  theories},''
\href{http://www.arXiv.org/abs/1708.02573}{{ 1708.02573}}.

\bibitem{Anastasiou:2017taf}
A.~Anastasiou, L.~Borsten, M.~J. Duff, A.~Marrani, S.~Nagy, and M.~Zoccali,
  ``{The Mile High Magic Pyramid},''
\href{http://www.arXiv.org/abs/1711.08476}{{ 1711.08476}}.
\newblock

\bibitem{Anastasiou:2018rd} 
  A.~Anastasiou, L.~Borsten, M.~J.~Duff, S.~Nagy and M.~Zoccali,
  ``Gravity as Gauge Theory Squared: A Ghost Story,''
  Phys.\ Rev.\ Lett.\  {\bf 121}, no. 21, 211601 (2018)
  [arXiv:1807.02486 [hep-th]].

\bibitem{Gurses:2018ckx} 
  M.~Gurses and B.~Tekin,
  ``Classical Double Copy: Kerr-Schild-Kundt metrics from Yang-Mills Theory,''
  Phys.\ Rev.\ D {\bf 98}, no. 12, 126017 (2018)
  [arXiv:1810.03411 [gr-qc]].

\bibitem{Anastasiou:2018rdx}
A.~Anastasiou, L.~Borsten, M.~J. Duff, S.~Nagy, and M.~Zoccali, ``{BRST
  squared},''
\href{http://www.arXiv.org/abs/1807.02486}{{ 1807.02486}}.

\bibitem{LopesCardoso:2018xes}
G.~Lopes~Cardoso, G.~Inverso, S.~Nagy, and S.~Nampuri, ``{Comments on the
  double copy construction for gravitational theories},'' in {\em {17th
  Hellenic School and Workshops on Elementary Particle Physics and Gravity
  (CORFU2017) Corfu, Greece, September 2-28, 2017}}.
\newblock 2018.
\newblock
\href{http://www.arXiv.org/abs/1803.07670}{{ 1803.07670}}.
\newblock

\bibitem{Goldberger:2019xef} 
  W.~D.~Goldberger and J.~Li,
  ``Strings, extended objects, and the classical double copy,''
  arXiv:1912.01650 [hep-th].

\bibitem{Lee:2018gxc} 
  K.~Lee,
  ``Kerr-Schild Double Field Theory and Classical Double Copy,''
  JHEP {\bf 1810}, 027 (2018)
  [arXiv:1807.08443 [hep-th]].

\bibitem{Cho:2019ype} 
  W.~Cho and K.~Lee,
  ``Heterotic Kerr-Schild Double Field Theory and Classical Double Copy,''
  JHEP {\bf 1907}, 030 (2019)
  [arXiv:1904.11650 [hep-th]].
    
\bibitem{Kim:2019jwm} 
  K.~Kim, K.~Lee, R.~Monteiro, I.~Nicholson and D.~Peinador Veiga,
  ``The Classical Double Copy of a Point Charge,''
  arXiv:1912.02177 [hep-th].
  
\bibitem{Luna:2015paa} 
  A.~Luna, R.~Monteiro, D.~O'Connell and C.~D.~White,
  ``The classical double copy for Taub-NUT spacetime,''
  Phys.\ Lett.\ B {\bf 750}, 272 (2015)
  [arXiv:1507.01869 [hep-th]].
  
    
\bibitem{Berman:2018hwd}
D.~S. Berman, E.~Chacon, A.~Luna, and C.~D. White, ``{The self-dual classical
  double copy, and the Eguchi-Hanson instanton},''
\href{http://www.arXiv.org/abs/1809.04063}{{ 1809.04063}}.
        

\bibitem{Huang:2019cja} 
  Y.~T.~Huang, U.~Kol and D.~O'Connell,
  ``The Double Copy of Electric-Magnetic Duality,''
  arXiv:1911.06318 [hep-th].
  

\bibitem{Alawadhi:2019urr}
  R.~Alawadhi, D.~S.~Berman, B.~Spence and D.~P.~Veiga,
  ``S-duality and the Double Copy,''
  arXiv:1911.06797 [hep-th].

\bibitem{Ehlers:1961zza}
  J.~Ehlers,
  ``Transformations of static exterior solutions of Einstein's gravitational field equations into different solutions by means of conformal mapping,''
  Colloq.\ Int.\ CNRS {\bf 91}, 275 (1962).

\bibitem{Geroch:1970nt} 
  R.~P.~Geroch,
  ``A Method for generating solutions of Einstein's equations,''
  J.\ Math.\ Phys.\  {\bf 12}, 918 (1971).

\bibitem{Harrison}
   B. K. Harrison, 
   ``New Solutions of Einstein-Maxwell Equations from Old,"
   J. Math. Phys. 9, 1744 (1968)

\bibitem{Maison:1979kx} 
  D.~Maison,
  ``Ehlers-harrison Type Transformations For Jordan's Extended Theory Of Gravitation,''
  Gen.\ Rel.\ Grav.\  {\bf 10}, 717 (1979).

\bibitem{Kinnersley:1977pg} 
  W.~Kinnersley,
  ``Symmetries of the Stationary Einstein-Maxwell Field Equations. 1.,''
  J.\ Math.\ Phys.\  {\bf 18}, 1529 (1977).

\bibitem{Kinnersley:1977ph} 
  W.~Kinnersley and D.~M.~Chitre,
  ``Symmetries of the Stationary Einstein-Maxwell Field Equations. 2.,''
  J.\ Math.\ Phys.\  {\bf 18}, 1538 (1977).

\bibitem{Galtsov:1995mb}
  D.~V.~Galtsov, A.~A.~Garcia and O.~V.~Kechkin,
  ``Symmetries of the stationary Einstein-Maxwell dilaton theory,''
  Class.\ Quant.\ Grav.\  {\bf 12} (1995) 2887
  [hep-th/9504155].

\bibitem{Chiodaroli:2014xia} 
  M.~Chiodaroli, M.~Günaydin, H.~Johansson and R.~Roiban,
  ``Scattering amplitudes in $ \mathcal{N}=2 $ Maxwell-Einstein and Yang-Mills/Einstein supergravity,''
  JHEP {\bf 1501}, 081 (2015)
  [arXiv:1408.0764 [hep-th]].

\bibitem{Chiodaroli:2015rdg} 
  M.~Chiodaroli, M.~Gunaydin, H.~Johansson and R.~Roiban,
  ``Spontaneously Broken Yang-Mills-Einstein Supergravities as Double Copies,''
  JHEP {\bf 1706}, 064 (2017)
  [arXiv:1511.01740 [hep-th]].
    
\bibitem{Chiodaroli:2016jqw} 
  M.~Chiodaroli,
  ``Simplifying amplitudes in Maxwell-Einstein and Yang-Mills-Einstein supergravities,''
  arXiv:1607.04129 [hep-th].

\bibitem{Goldberger:2016iau} 
  W.~D.~Goldberger and A.~K.~Ridgway,
  ``Radiation and the classical double copy for color charges,''
  Phys.\ Rev.\ D {\bf 95}, no. 12, 125010 (2017)
  [arXiv:1611.03493 [hep-th]].

\bibitem{Chester:2017vcz} 
  D.~Chester,
  ``Radiative double copy for Einstein-Yang-Mills theory,''
  Phys.\ Rev.\ D {\bf 97}, no. 8, 084025 (2018)
  [arXiv:1712.08684 [hep-th]].
    
\bibitem{Vaidya:1947zz} 
  P.~C.~Vaidya and P.~V.~Bhatt,
  ``A generalized Kerr-Schild metric,''
  Pramana {\bf 3}, 28 (1974).

\bibitem{Moynihan:2019bor} 
  N.~Moynihan,
  ``Kerr-Newman from Minimal Coupling,''
  JHEP {\bf 2001}, 014 (2020)
  [arXiv:1909.05217 [hep-th]].

\bibitem{White:2016jzc} 
  C.~D.~White,
  ``Exact solutions for the biadjoint scalar field,''
  Phys.\ Lett.\ B {\bf 763}, 365 (2016)
  [arXiv:1606.04724 [hep-th]].
         
\bibitem{Bakhmatov:2019dow}
  I.~Bakhmatov, N.~S.~Deger, E.~T.~Musaev, E.~\'O~Colg\'ain and M.~M.~Sheikh-Jabbari,
  ``Tri-vector deformations in $d=11$ supergravity,''
  JHEP {\bf 1908}, 126 (2019)
  [arXiv:1906.09052 [hep-th]].

\bibitem{Lunin:2005jy} 
  O.~Lunin and J.~M.~Maldacena,
  ``Deforming field theories with U(1) x U(1) global symmetry and their gravity duals,''
  JHEP {\bf 0505}, 033 (2005)
  [hep-th/0502086].

\bibitem{Buchdahl:1959nk} 
  H.~A.~Buchdahl,
  ``Reciprocal Static Metrics and Scalar Fields in the General Theory of Relativity,''
  Phys.\ Rev.\  {\bf 115}, 1325 (1959).
  
\bibitem{Momeni:2005uc} 
  D.~Momeni, M.~Nouri-Zonoz and R.~Ramezani-Arani,
  ``MM-NUT disk space via Ehlers transformation,''
  Phys.\ Rev.\ D {\bf 72}, 064023 (2005)
  [gr-qc/0508036].

\bibitem{Buchdahl:1959nk} 
  H.~A.~Buchdahl,
  ``Reciprocal Static Metrics and Scalar Fields in the General Theory of Relativity,''
  Phys.\ Rev.\  {\bf 115}, 1325 (1959).

\bibitem{Plebanski:1976gy}
J.~Plebanski and M.~Demianski,
``Rotating, charged, and uniformly accelerating mass in general relativity,''
Annals Phys. \textbf{98} (1976), 98-127

\bibitem{Bossard:2009at}
G.~Bossard, H.~Nicolai and K.~S.~Stelle,
JHEP \textbf{07}, 003 (2009)
[arXiv:0902.4438 [hep-th]].

\bibitem{Breitenlohner:1987dg}
P.~Breitenlohner, D.~Maison and G.~W.~Gibbons,
``Four-Dimensional Black Holes from Kaluza-Klein Theories,''
Commun. Math. Phys. \textbf{120}, 295 (1988)
             
\bibitem{Galtsov:2008zz} 
  D.~V.~Galtsov,
  ``Generating solutions via sigma-models,''
  Prog.\ Theor.\ Phys.\ Suppl.\  {\bf 172}, 121 (2008)
  [arXiv:0901.0098 [gr-qc]].

\bibitem{AlonsoAlberca:2000cs}
N.~Alonso-Alberca, P.~Meessen and T.~Ortin,
``Supersymmetry of topological Kerr-Newman-Taub-NUT-AdS space-times,''
Class. Quant. Grav. \textbf{17}, 2783-2798 (2000)
[arXiv:hep-th/0003071 [hep-th]].

\bibitem{Mann:2005mb} 
  R.~B.~Mann and C.~Stelea,
  ``New Taub-NUT-Reissner-Nordstrom spaces in higher dimensions,''
  Phys.\ Lett.\ B {\bf 632}, 537 (2006)
  [hep-th/0508186].

\bibitem{Chong:2004hw}
  Z.~W.~Chong, G.~W.~Gibbons, H.~Lu and C.~N.~Pope,
  ``Separability and killing tensors in Kerr-Taub-NUT-de sitter metrics in higher dimensions,''
  Phys.\ Lett.\ B {\bf 609} (2005) 124
  [hep-th/0405061].

\bibitem{Plebanski:1975xfb} 
  J.~F.~Pleba\~nski,
  ``A class of solutions of Einstein-Maxwell equations,''
  Annals Phys.\  {\bf 90}, no. 1, 196 (1975).

\bibitem{Emond:2020lwi}
W.~T.~Emond, Y.~T.~Huang, U.~Kol, N.~Moynihan and D.~O'Connell,
``Amplitudes from Coulomb to Kerr-Taub-NUT,''
[arXiv:2010.07861 [hep-th]].
  
\bibitem{Petkou:2015fvh} 
  A.~C.~Petkou, P.~M.~Petropoulos and K.~Siampos,
  ``Geroch group for Einstein spaces and holographic integrability,''
  PoS PLANCK {\bf 2015}, 104 (2015)
  [arXiv:1512.04970 [hep-th]].
  
\bibitem{Astorino:2012zm} 
  M.~Astorino,
  ``Charging axisymmetric space-times with cosmological constant,''
  JHEP {\bf 1206}, 086 (2012)
  [arXiv:1205.6998 [gr-qc]].
  
\end{thebibliography}
\end{document}